\documentclass{ieeeaccess}
\usepackage{cite}
\usepackage{graphicx}
\usepackage[bookmarks=false,hidelinks]{hyperref}
\usepackage{xurl}
\usepackage{booktabs}
\usepackage{array}
\usepackage{pifont}
\usepackage{listings}
\providecommand{\xfigwd}{0pt}

\makeatletter

\newif\if@biographyTOCentrynotmade
\global\@biographyTOCentrynotmadetrue
\makeatother

\newcommand{\yes}{\ding{51}}
\newcommand{\no}{\ding{55}}
\newcommand{\pmark}{$\sim$}

\begin{document}
\history{Date of publication xxxx 00, 0000, date of current version xxxx 00, 0000.}
\doi{10.1109/ACCESS.2017.DOI}

\title{Enhancing RAMOSE, a Framework for Implementing REST APIs and Semantic-Actionable Outputs Over Data Sources}
\author{
\uppercase{Arcangelo Massari}\authorrefmark{1},
\uppercase{Sergei Slinkin}\authorrefmark{2},
\uppercase{Ivan Heibi}\authorrefmark{1},
\uppercase{and Silvio Peroni}\authorrefmark{1}
}

\address[1]{Research Centre for Open Scholarly Metadata, Department of Classical Philology and Italian Studies, University of Bologna, 40126 Bologna, Italy (e-mail: \{arcangelo.massari, ivan.heibi2, silvio.peroni\}@unibo.it)}
\address[2]{Digital Humanities and Digital Knowledge, Department of Classical Philology and Italian Studies, University of Bologna, 40126 Bologna, Italy (e-mail: sergei.slinkin@studio.unibo.it)}

\tfootnote{This work was funded by the European Union's Horizon Europe programme, Grant Agreement No.\ 101188018 (GRAPHIA).}

\markboth
{A. Massari \headeretal: Enhancing RAMOSE}
{A. Massari \headeretal: Enhancing RAMOSE}

\corresp{Corresponding author: Arcangelo Massari (e-mail: arcangelo.massari@unibo.it).}

\begin{abstract}
Scholarly infrastructures increasingly expose their data through REST APIs that follow shared specifications, such as the Scientific Knowledge Graphs - Interoperability Framework (SKG-IF), which defines a common data model, exchange format, and REST API for research information. Implementing such specifications over existing data sources, however, requires a development effort that many open infrastructures cannot afford. RAMOSE, the RESTful API Manager Over SPARQL Endpoints, is an open-source Python framework that reduces this effort by turning a declarative configuration file into a documented REST API over RDF triplestores. This article presents its second major version, which addresses nine new requirements. The new features include query orchestration across multiple SPARQL endpoints and non-RDF sources, with joins across their results; pluggable output formats and request parameters; pagination and caching; OpenAPI export; and write operations, with authentication for both API consumers and protected endpoints. A built-in module packages the format and filters that SKG-IF prescribes, letting a provider expose a compliant endpoint only through configuration. A functional comparison with nine similar tools, grounded in reproducible tests, shows that RAMOSE is the only one able to serve both RDF and non-RDF sources simultaneously within a single API operation, joining their results on arbitrary keys. Its first version served the OpenCitations REST APIs, which peaked at almost 38 million monthly requests between May 2025 and May 2026. The new version extends this deployment to the OpenCitations SKG-IF endpoint and is adopted by the GRAPHIA project to onboard data sources into its SKG-IF-based federation.
\end{abstract}


\titlepgskip=-15pt

\maketitle

\section{Introduction}
\label{sec:introduction}

% principle of open science about interoperability and, as related, fairness
% unesco_unesco_2021 wilkinson_fair_2016 posi_adopters_principles_2025 corcho_maturity_2024
\PARstart{I}{nteroperability} is a concept that has characterised several aspects of the Open Science construct defined by UNESCO~\cite{unesco_unesco_2021}, and it is usually implemented, e.g. in the \textit{European Open Science Cloud} (EOSC, \url{https://eosc.eu/})~\cite{burgelman_politics_2021}, by adopting shared formal conceptualisation such as vocabularies, taxonomies, ontologies, or, more generally, \textit{semantic artefacts}~\cite{corcho_eosc_2021,corcho_maturity_2024,nyberg_akerstrom_developing_2024}. Using these artefacts is crucial for scholarly infrastructures~\cite{posi_adopters_principles_2025} that aim to interoperate in a network of distributed services and comply with the FAIR principles~\cite{wilkinson_fair_2016}.

The various strategies often adopted to guarantee that different systems included in a distributed environment, with their own workflows and data, can interoperate with each other include the use of REST APIs~\cite{chue_hong_fair_2022}, which are usually ``natively supported by all major programming and runtime frameworks today''~\cite{kakaletris_design_2023}, even if they should be chosen as a solution and implemented to be compliant with technical standards to maximise interoperability and reuse of code~\cite{scardaci_technical_2025,baumann_landscape_2026}. For instance, in this context, a recent standard developed to guarantee the interoperability of infrastructures sharing data about research (bibliographic metadata, citation data, funding information, etc.) is the \textit{Scientific Knowledge Graphs - Interoperability Framework} (SKG-IF, \url{https://skg-if.github.io/}), which provides a data model for describing its data, a JSON-LD format for exchanging them, and a REST API specification detailing basic operations for retrieving SKG-IF data in JSON-LD format \cite{mannocci_scientific_2025}.

While standard API definitions exist, such as SKG-IF, putting them into practice, and thus implementing them in existing systems, requires significant effort, including both the human time needed to provide them and meaningful financial contributions to support development and maintenance. Often, existing open scholarly infrastructures and systems struggle with financial sustainability, and even if implementing new API endpoints could increase their visibility, interoperability and data reuse, the costs are often beyond their capacity. Therefore, wider implementation of REST APIs, particularly when they should follow formalised and shared specifications, also depends on services that reduce the time and financial costs of adopting such standards.

In 2018, OpenCitations (\url{https://opencitations.net})~\cite{peroni_opencitations_2020}, developed and released a software tool called RAMOSE, the RESTful API Manager Over SPARQL Endpoints (\url{https://github.com/opencitations/ramose/})~\cite{daquino_creating_2022, massari_ramose_2026} to simplify the creation of REST APIs for RDF, which is the general data model used by OpenCitations to store its data. RAMOSE is a Python software package for creating a Web REST API, with related documentation, that serves as an interface to a SPARQL endpoint, regardless of the data types hosted in the RDF triplestore. The REST API is created by developing a semi-structured textual configuration file (which includes the SPARQL queries the API uses to retrieve RDF data) plus a few Python files if add-on functions for data pre- and post-processing are needed. While RAMOSE has been adopted in several use cases and is used, in particular, to deliver all the OpenCitations REST APIs (\url{https://api.opencitations.net}), thereby providing a configurable mechanism to enable initial technical interoperability, it lacked several features that the community had strongly requested.

In this article, we present the result of the development work we have performed in the past years on RAMOSE to address these requests, which includes gathering data resulting from a REST API operation from multiple SPARQL and non-SPARQL sources, joining the results of several queries into a single response, customising the API parameters and return formats, caching and paging the results, providing REST API documentation following shared standards (i.e. OpenAPI), and going beyond the data reading activity thus enabling the REST APIs also for writing tasks.

The rest of the article is organised as follows. In Section~\ref{sec:requirements}, we introduce the requirements for the development of RAMOSE, including both the original ones devised in 2018 and the new ones that characterise this new version. In Section~\ref{sec:ramose}, we present the overall architecture of RAMOSE, focusing on the features implemented to address the new requirements. Section~\ref{sec:uptake}, instead, describes how the community have adopted RAMOSE, both in terms of existing instances and its adoption in specific projects for addressing technical and semantic interoperability problems, while Section~\ref{sec:related} presents related tools developed in the past for addressing similar issues, adding a functional comparative analysis between them and RAMOSE. Section~\ref{sec:evaluation} measures the query orchestration of RAMOSE against a federated SPARQL query on OpenCitations data. Finally, in Section~\ref{sec:conclusions}, we conclude the article, outlining some possible future developments.

\section{Original and new requirements}
\label{sec:requirements}

The design of the first version of RAMOSE rested on seven requirements~\cite{daquino_creating_2022}, which remain the starting point for the new version:

\begin{enumerate}
    \item It must work with any RDF triplestore providing a SPARQL endpoint.
    \item A Semantic Web expert should only be required to define the SPARQL queries behind the API operations, while all the other aspects of the REST API configuration and use should not require Semantic Web skills.
    \item API operations and their input parameters must be fully customisable.
    \item The configuration file must be easy to write and must avoid technicalities as much as possible.
    \item It must be possible to specify pre-processing and post-processing steps in any operation, developed as pure Python functions, so as to better customise the interpretation of the input parameters and call outputs.
    \item Basic built-in filters and refinement mechanisms must be provided by default.
    \item It must be possible to use the REST API within another Python application, to run it as a command-line application, and to make it available as a service within a web server.
\end{enumerate}

These original requirements were drawn from the goal of exposing OpenCitations RDF data. The new requirements address the needs of SKG-IF, the interoperability framework introduced in Section~\ref{sec:introduction}.

An SKG-IF entity is a record describing a scholarly object and links to related entities. For example, a research product may carry bibliographic metadata and links to the products it cites. These two kinds of data may live in separate datasets: at OpenCitations, for instance, the metadata are in OpenCitations Meta~\cite{massari_opencitations_2024} and the citations in OpenCitations Index~\cite{heibi_opencitations_2024}, each behind its own SPARQL endpoint, so assembling one entity requires querying both and joining their results. A single federated SPARQL query with \texttt{SERVICE} clauses~\cite{prudhommeaux_sparql_2013} is the native SPARQL way to do this. When the query mixes local patterns with a \texttt{SERVICE} block, the engine can first find matching items in its own data and then send them to the remote endpoint to retrieve related data. The Recommendation lets the engine send these items one at a time or in groups, and each group costs one HTTP request. A query that matches thousands of items can thus cause thousands of remote requests, and their number is a known cost in studies of federation systems~\cite{saleem_fine-grained_2016}. This has three effects for a provider. (1) Every remote request can end in a timeout or in throttling, because the remote endpoint decides how much load it accepts. A query with many matching items therefore takes minutes even when each remote request is fast, as the measurement in Section~\ref{sec:evaluation} shows. (2) The query is all-or-nothing, because sub-queries cannot run on their own, so one failing block fails the whole request and cannot be retried in isolation. Correctness and speed thus depend on systems outside the provider's control. (3) Some endpoints restrict federation because it threatens their stability: the Wikidata Query Service, for example, allows federation only from an allowlist (\url{https://www.wikidata.org/wiki/Wikidata:SPARQL_query_service/Federated_queries/Allowlist}). Issuing the queries separately and joining their results afterwards solves these problems, because the provider decides how many bindings each request carries and can retry each step on its own.

The integration that SKG-IF targets is even broader, as it is meant to bridge sources that adopt different data models and technologies, not all of them RDF. This follows the broad, model-agnostic notion of knowledge graph that SKG-IF adopts, namely ``a graph of data intended to accumulate and convey knowledge of the real world, whose nodes represent entities of interest and whose edges represent potentially different relations between these entities''~\cite{hogan_knowledge_2022}. The same mechanism must therefore be able to read non-RDF sources.

SKG-IF also prescribes how the data must be represented. Its entities are exchanged as JSON-LD, which neither the CSV nor the generic JSON of the first version of RAMOSE can produce. In addition, its search operations accept a defined filter vocabulary. The API must therefore be able to return that format and to expose the parameters its filter vocabulary requires. SKG-IF search operations return paged results, so a conforming API must paginate, and caching responses keeps paging efficient and lightens the load on the endpoints. 

Then, SKG-IF publishes its own contract as an OpenAPI specification~\cite{openapi_initiative_openapi_2025}, so providers can describe their API in the same machine-readable form. Finally, since SKG-IF is not specific to OpenCitations but a standard adopted by a broad community, packaging this support as a built-in, reusable module would lower the barrier to publishing data through SKG-IF and thereby widen the network of providers that expose their data in an interoperable way.

From these needs, we derived six new RAMOSE requirements:

\begin{description}
    \item[\textbf{REQ1}]: Gather the data for an operation from multiple sources, including non-RDF ones.
    \item[\textbf{REQ2}]: Run several queries within one operation and join their results.
    \item[\textbf{REQ3}]: The API designer can plug in a custom output format and custom REST parameters for an operation.
    \item[\textbf{REQ4}]: Paginate the responses of an operation and cache them.
    \item[\textbf{REQ5}]: Export the API description as an OpenAPI specification.
    \item[\textbf{REQ6}]: Package reusable support for SKG-IF as a built-in module that supplies the format and filters the standard requires.
\end{description}

Three further requirements do not descend from SKG-IF but from recurring requests of the community running RAMOSE. The first version served read-only APIs. The community asked to lift that limit, so that an operation may also modify the data behind the endpoint. This ability sets up the other two. Once an operation can write, leaving it open to anyone on the Web is untenable, so it must be possible to confine such an operation to authenticated clients. The data source poses a matching condition from the opposite side: it may respond only to a caller that presents a credential, as QLever~\cite{bast_qlever_2017} does by default for the operations that modify its data. Hence the seventh, eighth, and ninth requirements:

\begin{description}
    \item[\textbf{REQ7}]: Operations can modify the data behind the endpoint.
    \item[\textbf{REQ8}]: Restrict a data-modifying operation to authenticated clients.
    \item[\textbf{REQ9}]: Enable interaction with a data source that requires the caller to present a credential.
\end{description}

\section{RAMOSE: architecture and new features}
\label{sec:ramose}

RAMOSE is a middleware layer interposed between an HTTP client and one or more data sources, turning a declarative configuration into a documented REST API whose operations are backed by SPARQL queries. The version presented here preserves this model and the hash-format configuration file (extension \texttt{.hf}) introduced in the first version, i.e. a Markdown-based syntax in which each field is a hashtag-prefixed key followed by its value. It also accepts YAML files as an alternative syntax whose fields mirror the hash format. Backward compatibility is preserved: operations written in the original single-query syntax run unchanged, and the new behaviour is activated only when the new fields and directives described below are used. The remainder of this section presents the new features, relating each to the requirements of Section~\ref{sec:requirements}.

\subsection{Code development and organisation}
\label{sec:development}

The first version of RAMOSE was distributed as a single Python script. RAMOSE is now a Python package published on PyPI~\cite{massari_ramose_2026}. Its components are organised by responsibility, as shown in Figure~\ref{fig:architecture}: an \texttt{APIManager} loads the API specification and routes each request to an \texttt{Operation}, which runs the processing pipeline, invoking an optional addon module and the result cache, and querying SPARQL endpoints directly or non-RDF sources through SPARQL Anything~\cite{daga_facadex_2021}. Before dispatching a request to an operation that requires authentication, the \texttt{APIManager} checks the caller's credentials against a token store.

\Figure[!htb][width=0.92\textwidth]{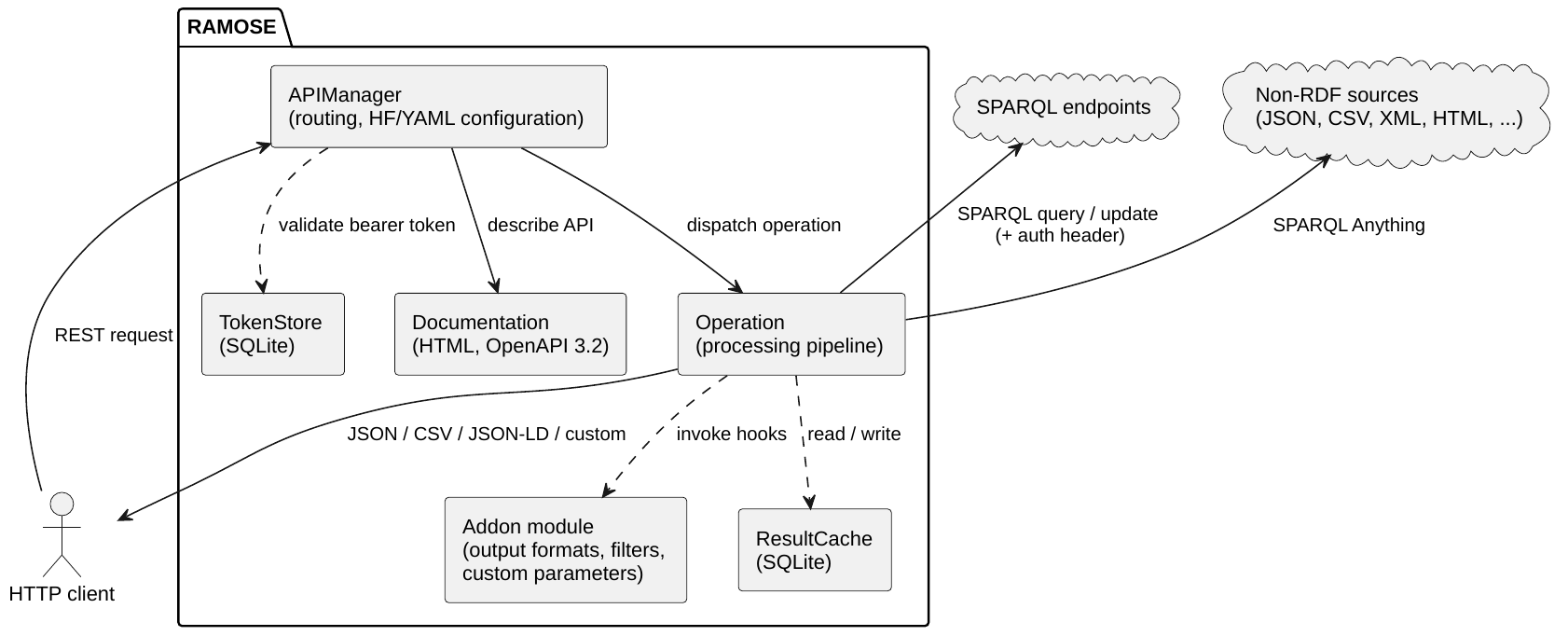}{The main components of RAMOSE and the data sources it mediates, with REST requests entering from the HTTP client.\label{fig:architecture}}

As for the development methodology, test-driven development~\cite{beck_testdriven_2003} guided the implementation. Beyond unit tests, the suite includes integration tests that exercise RAMOSE against a triplestore. A continuous integration configuration~\cite{humble_continuous_2011} runs the full suite on every commit through GitHub Actions across multiple Python versions and reports test coverage, which stands at 94\% at the time of writing; separate workflows enforce linting and static type checking.

Releases follow semantic versioning (\url{https://semver.org/}), and the changelog is generated from Conventional Commits (\url{https://www.conventionalcommits.org/en/v1.0.0/}). Per-file licensing is made machine-readable through the REUSE specification (\url{https://reuse.software/spec-3.3/}) and verified in continuous integration. Finally, the package is accompanied by a documentation site (\url{https://opencitations.github.io/ramose/}).

\subsection{Request processing workflow}
\label{sec:workflow}

Figure~\ref{fig:workflow} traces how an operation is served. RAMOSE first matches the request URL to an operation; if that operation is protected, the caller's token is checked immediately, and a missing or invalid one ends the exchange with a 401 Unauthorized response before any further work. Otherwise, RAMOSE extracts the parameters, converts each to the data type declared in the configuration, and optional preprocessing functions may adjust those values before substituting them into the SPARQL query. A data-modifying operation then takes a path of its own: it runs its update against the endpoint, empties the cache, whose entries would otherwise still reflect the data as it was before the change, and returns, without entering the read steps that follow. For a read operation, the result is sought in the cache, and only on a miss is the query actually executed: against a single SPARQL endpoint, against a non-RDF source through SPARQL Anything, or, when the query carries the multi-source directives introduced below, through an orchestration loop that assembles the answer step by step into an accumulator before re-entering the common path. For each field in the retrieved rows, RAMOSE keeps the text value alongside a typed value based on the API configuration. Both values are available to optional post-processing functions, after which the built-in \texttt{require}, \texttt{filter}, and \texttt{sort} parameters refine the results. Comparisons and sorting use the typed values. RAMOSE then removes the typed values and caches the table of text values. Finally, if the request specifies it, RAMOSE paginates the response and serialises it into the requested format.

\Figure[!htb][width=0.92\textwidth]{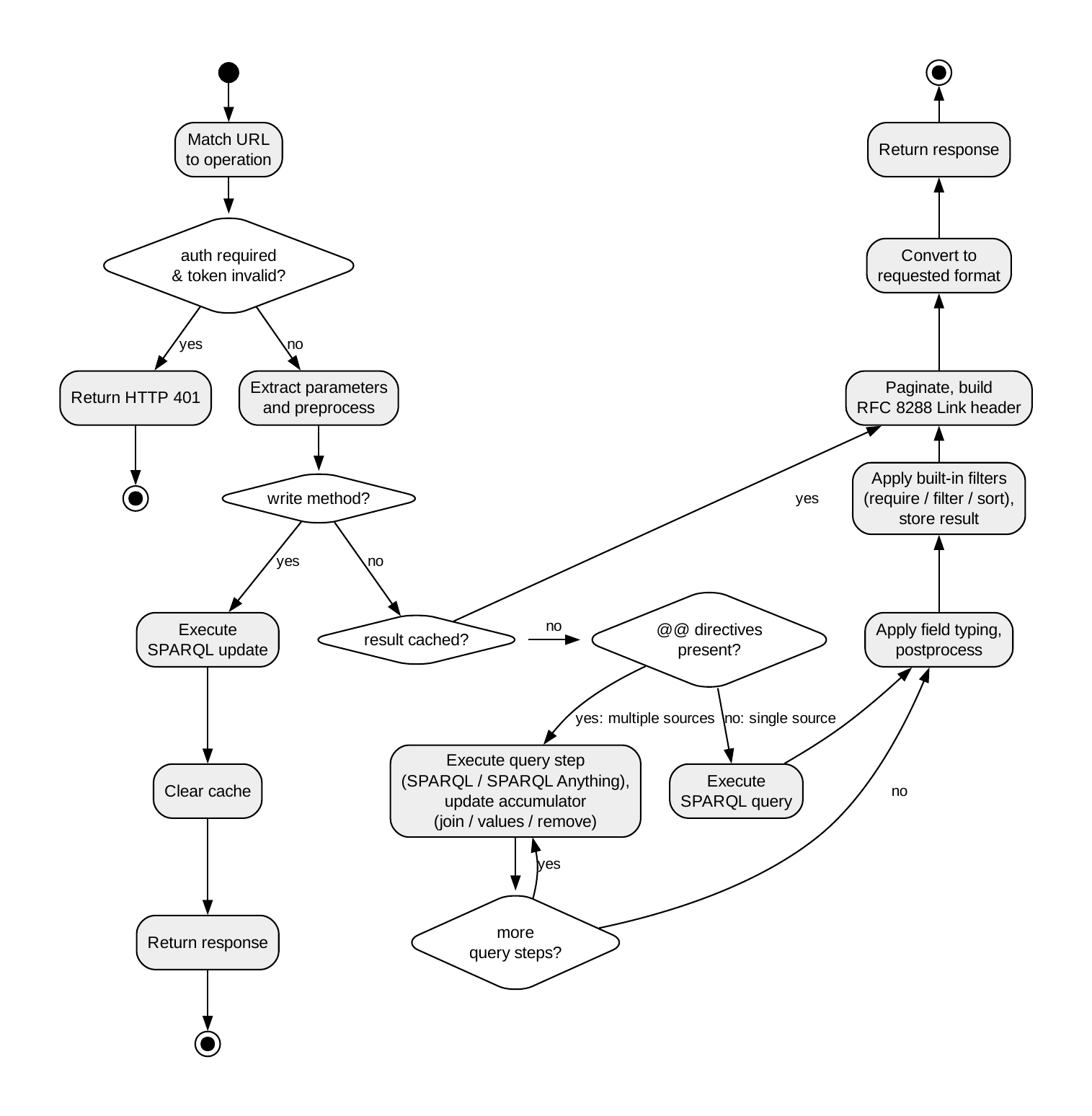}{The workflow RAMOSE follows from an operation request to the response returned.\label{fig:workflow}}

\subsection{Query orchestration: multiple sources and joins}
\label{sec:multisource}

The first two new requirements (\textit{REQ1} and \textit{REQ2}) are handled together. Whereas the first version bound an API to a single endpoint, the HF or YAML configuration file that defines an API can now list several: its \texttt{sources} field maps a short name to each endpoint URL, as in \texttt{meta=}\url{https://sparql.opencitations.net/meta}, and a query step chooses where it runs with the \texttt{@@with} directive, naming one of those sources (\texttt{@@with source=meta}) or giving an endpoint URL directly. The same directive also picks the backend: by default the step is sent to a SPARQL endpoint over HTTP, while \texttt{@@with engine=sparql-anything} reads non-RDF inputs as virtual RDF~\cite{daga_facadex_2021} through the PySPARQL-Anything library~\cite{ratta_pysparqlanything_2026}. At the time of writing, the latter covers XML, JSON, CSV, HTML, Excel, TXT, binary (non-text files such as images), EXIF, Zip, Tar, Markdown, YAML, BibTeX, DOCX, and PPTX. Either way, the output is normalised into the same tabular rows, so the steps that follow are agnostic to where the data came from.

An operation is thus written as a sequence of SPARQL queries interleaved with \texttt{@@} directives that share an accumulator holding the rows produced so far. Listing~\ref{lst:multisource} is a complete operation of this kind. Given the ISSN of a venue, it returns the venue's articles, their incoming citations, and the title and date of each citing work. The articles and the metadata come from OpenCitations Meta, while the citations come from OpenCitations Index. Its first query runs against the default Meta endpoint and maps the ISSN to the OpenCitations Meta Identifiers (OMIDs) of the articles, bound to \texttt{?omid}. Each later step adds a query whose result is added to the accumulator by \texttt{@@join}. Its two arguments name the join key in the accumulator and in the next query result. Its \texttt{type} keeps only the rows found on both sides (\texttt{inner}) or every accumulated row, even when the second query has no match for it (a \texttt{left} outer join). Both joins in the operation use \texttt{type=left}, so an article that Index does not cite is still returned, with its citation columns left empty.

RAMOSE retains the CSV tabular model of its first version and compares join keys from RDF and non-RDF sources as strings. It uses the bare text of an IRI or the lexical form of a literal, ignoring datatypes and language tags. This is consistent with the W3C specification for SPARQL results in CSV, which omits RDF term types, datatypes, and language tags~\cite{seaborne_sparql_csv_2013}. When datatype or language distinctions matter, API authors can encode the datatype or language tag in the key string within each query. Keys that are missing or empty after trimming do not match, including empty literals represented as empty CSV cells. When a key occurs in several rows on either side, each accumulated row is paired with every matching row.

The second step selects the Index endpoint with \texttt{@@with source=index} and feeds it the OMIDs already gathered: \texttt{@@values ?omid} reads the listed variable from the accumulator and writes its values into the next query as a SPARQL \texttt{VALUES} clause, so Index is asked only about the articles found in Meta. The third step goes back to Meta with the citing works. Since a venue may have tens of thousands of them, the optional \texttt{batch\_size} argument splits the values into blocks of the given size, and the query runs once per block, so as not to overload the endpoint.

When a transient error occurs, RAMOSE retries the failed request and keeps the results of earlier steps during the same execution. One can configure the number of attempts and waiting intervals globally or per operation. If a step fails and cannot be retried further, the operation returns an error without partial results.

\begin{lstlisting}[float=*,caption={A multi-source operation joining the articles of a venue from OpenCitations Meta with their citations from OpenCitations Index.},label={lst:multisource}]
#url /api/v1
#type api
#endpoint https://sparql.opencitations.net/meta
#sources meta=https://sparql.opencitations.net/meta; index=https://sparql.opencitations.net/index

#url /venue-citations/{issn}
#type operation
#issn str(.+)
#method get
#sparql PREFIX datacite: <http://purl.org/spar/datacite/>
PREFIX literal: <http://www.essepuntato.it/2010/06/literalreification/>
PREFIX frbr: <http://purl.org/vocab/frbr/core#>
PREFIX fabio: <http://purl.org/spar/fabio/>
SELECT DISTINCT ?omid WHERE {
  ?id datacite:usesIdentifierScheme datacite:issn ;
      literal:hasLiteralValue "[[issn]]" .
  ?venue datacite:hasIdentifier ?id .
  ?omid frbr:partOf+ ?venue ; a fabio:JournalArticle .
}

@@with source=index
@@join left_var=?omid right_var=?omid type=left
@@values ?omid
PREFIX cito: <http://purl.org/spar/cito/>
SELECT DISTINCT ?omid ?citation ?citing WHERE {
  ?citation a cito:Citation ; cito:hasCitedEntity ?omid ;
            cito:hasCitingEntity ?citing .
}

@@with source=meta
@@join left_var=?citing right_var=?citing type=left
@@values ?citing batch_size=3000
PREFIX dcterms: <http://purl.org/dc/terms/>
PREFIX prism: <http://prismstandard.org/namespaces/basic/2.0/>
SELECT ?citing ?title ?date WHERE {
  ?citing dcterms:title ?title .
  OPTIONAL { ?citing prism:publicationDate ?date }
}
\end{lstlisting}

\subsection{Custom output formats}
\label{sec:output}

The first version of RAMOSE could return CSV and a JSON rendering derived from the same flat table. RAMOSE v2 expands this model: each operation can declare additional output formats. A \texttt{\#format} field maps a public format name to a function provided by the operation's addon module; an optional third value declares the media type.

The conversion happens at the end of the processing pipeline. RAMOSE still normalises the query output into the same internal table used by CSV and JSON, applies the configured processing steps, and only then serialises the result. A converter receives that table as a CSV string. If a media type is declared for the format, RAMOSE can also use it in HTTP content negotiation and in the generated OpenAPI response schema. The converter may emit XML, RDF serialisations, JSON-LD, or an application-specific structure without changing the query execution layer.

\subsection{Custom request parameters}
\label{sec:custom-params}

RAMOSE v1 already had built-in query parameters for table-level operations such as requiring non-empty fields, filtering, sorting, selecting CSV or JSON output, and reshaping JSON rows. RAMOSE v2 keeps these parameters, but also lets an operation define its own request parameters (\textit{REQ3}). This is needed, for instance, when an API must expose a domain-specific filter vocabulary, or when a request parameter has to change the query before it is executed. Query-time handling can be required for performance: a title-search filter can use a database text index, whereas a row filter would first retrieve the whole table and only then discard non-matching rows.

The \texttt{\#custom\_params} field lets an operation replace or add query parameters, and the behaviour of a custom parameter can be implemented in two ways: procedurally, by writing a Python function, or declaratively, by describing the parameter in a YAML configuration file. The procedural mode gives full control: each entry names the public parameter, the function that handles it, the processing phase, and the documentation text. The function may run in the \texttt{preprocess} phase, where it receives the request values and returns a mapping from placeholder names to SPARQL fragments, or in the \texttt{postprocess} phase, where it receives the plain result table and can transform it after the built-in filters. When a custom parameter has the same name as a built-in one, the built-in behaviour is disabled for that operation. This became necessary because SKG-IF defines a \texttt{filter} parameter with its own logic and semantics.

The declarative mode reduces configuration effort when a parameter only needs to inject a fixed SPARQL fragment, sparing the developer from writing any code. Here the parameter is described in a YAML file that maps each accepted filter key to one or more named slots, and each slot contains the SPARQL template that will be injected into the matching \texttt{[[...]]} placeholder in the operation query. The placeholder \texttt{\{\{value\}\}} is replaced with the value supplied by the client. The YAML configuration can also inject the \texttt{@@} directives described in Section~\ref{sec:multisource}, so a query parameter can activate a join with another declared source when the filter needs data held elsewhere. This declarative mode was introduced to ease the adoption of SKG-IF.

To prevent SPARQL injection, RAMOSE escapes the textual values that clients provide before adding them to a query. It also casts numbers and dates to their declared types and rejects IRI values if they contain forbidden characters. In a YAML template, RAMOSE escapes the value supplied for \texttt{\{\{value\}\}} as a string literal when the placeholder is enclosed in quotes, and checks it for forbidden IRI characters when enclosed in angle brackets.

\subsection{Paging and caching}
\label{sec:caching}
To handle \textit{REQ4}, RAMOSE offers pagination at three levels. The simplest is built in and needs no configuration: a request carrying \texttt{page} and \texttt{page\_size} parameters receives the corresponding window of result rows. RAMOSE slices its internal table to that window and adds an RFC 8288~\cite{nottingham_web_2017} \texttt{Link} header whose \texttt{first}, \texttt{prev}, \texttt{next}, and \texttt{last} relations carry the URLs of the neighbouring pages. Here the total number of items is the row count, which is correct only when one result row corresponds to one item.

This row-based scheme is the default precisely because RAMOSE is agnostic to the data model: knowing nothing about how the underlying entities are shaped, it treats the result row as the only unit it can count and slice. When the output groups several rows into one entity, however, the row ceases to be the natural unit, and a row-based window could split an entity across two pages. For these cases RAMOSE lets pagination be made entity-aware, driven by the format itself. The converter described in Section~\ref{sec:output} receives the absolute request URL together with the result table. Then it can read the \texttt{page} and \texttt{page\_size} parameters from that URL to paginate by entity rather than by row, counting the entities, validating the requested page, cutting it out, and embedding custom navigation metadata in its own output structure.

Both schemes still compute the entire result set before discarding all but one page. For large collections, this is wasteful: an SKG-IF \texttt{products} endpoint that resolves the full tree of each work---its authors, venues, identifiers, and citations---would build that tree for every match only to return ten. The \texttt{@@page} directive avoids this. The directive takes the query variable bound to the entities to be paginated and keeps only those of its distinct values that fall on the requested page, discarding the other rows. The intended use is to place it just after a cheap first query that retrieves only those entities, and before the queries that resolve their full data, so that the expensive resolution runs for one page of entities rather than for the whole match set. Listing~\ref{lst:page} shows this pattern in the SKG-IF \texttt{products} endpoint, where the named \texttt{default\_size} and \texttt{max\_size} arguments set and bound the page size.

\begin{lstlisting}[float=*,caption={The \texttt{@@page} directive in the SKG-IF \texttt{products} operation.},label={lst:page}]
#url /products
#type operation
#method get
#custom_params filter,skgif_filters.ocdm.yaml,Search filter.

#sparql [[filter_preamble]]
@@with source=meta
PREFIX fabio: <http://purl.org/spar/fabio/>
SELECT DISTINCT ?local_identifier WHERE {
  ?local_identifier a fabio:Expression .
  [[filter]]
}
ORDER BY ?local_identifier

@@page ?local_identifier default_size=10 max_size=100

@@values ?local_identifier
@@join left_var=?local_identifier right_var=?local_identifier type=left
@@with source=meta
PREFIX dcterm: <http://purl.org/dc/terms/>
SELECT ?local_identifier ?title WHERE {
  ?local_identifier dcterm:title ?title .
}
\end{lstlisting}

Caching addresses the complementary cost of recomputing identical requests. RAMOSE stores each operation's result in a SQLite database, keyed by the endpoint, the operation URL, and every request parameter that changes the table, namely the path parameters, the built-in \texttt{require}, \texttt{filter}, and \texttt{sort}, and the custom parameters. What is cached is the post-processed table, after those filters and custom parameters have been applied but before format conversion: both the built-in and the format-driven pagination act on that table only at conversion time, so they are purely presentational, and the format and the page are left out of the key, letting one cached result serve every format and every page of the same query. The \texttt{@@page} directive is the exception, because it paginates during query execution: it changes which entities are fetched and thus the table itself, so when an operation uses it the page parameters become part of the key, and each page is cached on its own. Each entry carries an expiry; the default lifetime is one day. An operation can shorten or lengthen it with \texttt{\#cache\_duration} or opt out entirely with \texttt{\#cache\_disable}. Finally, a successful write operation clears the cache so that later reads do not return stale data.

\subsection{OpenAPI export}
\label{sec:openapi}

To handle \textit{REQ5}, RAMOSE now generates an OpenAPI~3.2 specification using the same \texttt{.hf} configuration that defines the operations. What makes the export more than a transcription is that the specification reflects the configuration choices described above rather than a fixed template. The custom output formats of Section~\ref{sec:output} appear as response content under the media type each declares, such as \texttt{application/ld+json} for SKG-IF. Parameters removed with \texttt{disable\_params} disappear from the contract, and the custom parameters of Section~\ref{sec:custom-params} are listed in their place. For protected operations, whose authentication mechanism is described in Section~\ref{sec:auth}, the specification also includes a bearer security scheme and a 401 response.

The schema of each response is taken from the declared field types or, when none are given, inferred from the example output recorded in the configuration, while the directives that drive RAMOSE internally, such as the SPARQL queries, the pre and post-processing functions, and the addon reference, are deliberately left out, being internal implementation details of no interest to an external consumer. RAMOSE serves the OpenAPI specification as a YAML document, and renders it as a Swagger UI page at the \texttt{/docs} route for human browsing.

\subsection{SKG-IF support module}
\label{sec:skgif}

SKG-IF is an interoperability framework, as introduced in Section~\ref{sec:introduction}. Its value grows with the number of endpoints that expose data according to its data model. To support that direction, RAMOSE includes a built-in module for SKG-IF. The module allows providers to publish query results as SKG-IF JSON-LD and expose the SKG-IF filter parameter through declarative configuration, without writing code.

The SKG-IF module defines a contract for result column names, which come from the variables selected in SPARQL. Providers only need to write SPARQL queries and filter patterns over their own data model, using those names for the columns. To convert the query results, the module groups rows that refer to the same entity by their local identifier. It then places their values into the nested objects and lists defined by the SKG-IF specification. The module also adds metadata identifying the response by its URL and type. For searches, the metadata identify the full result set, report the total result count and link to the first and last pages, plus previous and next pages when available.

Listing~\ref{lst:skgif-output-filter} gives an example of this configuration. It also shows how the custom formats of Section~\ref{sec:output} and custom parameters of Section~\ref{sec:custom-params} can be used together. The API specification imports \texttt{ramose.skg\_if}; the operation registers the SKG-IF output format, makes it the default response, and binds the SKG-IF \texttt{filter} parameter to a YAML configuration file.

\begin{lstlisting}[float=*,caption={Combining a custom output format and a config-driven custom parameter for an SKG-IF operation.},label={lst:skgif-output-filter}]
#url /skg-if/v1
#type api
#base https://api.opencitations.net/skg-if
#endpoint https://sparql.opencitations.net/meta
#sources meta=https://sparql.opencitations.net/meta; index=https://sparql.opencitations.net/index
#addon ramose.skg_if
#disable_params require,filter,sort,format,json

#url /products
#type operation
#method get
#description Returns a list of research products matching the given filters.
#custom_params filter,skgif_filters.ocdm.yaml,Search filter.
#format skg_if,to_skg_if,application/ld+json
#default_format skg_if
#sparql [[filter_preamble]]
PREFIX dcterm: <http://purl.org/dc/terms/>

SELECT ?local_identifier ?title WHERE {
  ?local_identifier dcterm:title ?title .
  [[filter]]
}
\end{lstlisting}

Listing~\ref{lst:skgif-yaml-filter} shows a title filter from the YAML file used by the operation in Listing~\ref{lst:skgif-output-filter}.

\begin{lstlisting}[basicstyle=\ttfamily\scriptsize,caption={YAML configuration of an SKG-IF title filter.},label={lst:skgif-yaml-filter}]
cf.search.title:
  filter: |-
    ?local_identifier dcterm:title ?title .
    FILTER(CONTAINS(?title, "{{value}}"))
\end{lstlisting}

The request in Listing~\ref{lst:skgif-title-request} searches for products whose title contains \texttt{OpenCitations}.

\begin{lstlisting}[basicstyle=\ttfamily\scriptsize,caption={SKG-IF request to search for products by title.},label={lst:skgif-title-request}]
GET https://api.opencitations.net/skg-if/v1/products
    ?filter=cf.search.title:OpenCitations
\end{lstlisting}

The request fills \texttt{\{\{value\}\}} with \texttt{OpenCitations}; RAMOSE then places the rendered SPARQL fragment in the operation query. The same YAML slot can hold a database-specific text-index pattern; in Virtuoso, for example, it can use \texttt{bif:contains} on \texttt{?title}.

\subsection{Write operations and authentication}
\label{sec:auth}

These last three requirements (\textit{REQ7}, \textit{REQ8} and \textit{REQ9}) extend RAMOSE from a tool confined to reading open endpoints into one that can modify a source, guard the operations that do so, and authenticate to a source that requires it. \textit{REQ7} is enabled by declaring an HTTP write method, \texttt{post}, \texttt{put}, or \texttt{delete}, in an operation's \texttt{\#method} field; its \texttt{\#sparql} then carries a SPARQL~1.1 Update~\cite{gearon_sparql_2013} instead of a query. Update parameters follow the same rules as query parameters (Section~\ref{sec:custom-params}). RAMOSE sends the update to the endpoint named in \texttt{\#update\_endpoint}, defaulting to the read endpoint when that field is absent, and on success clears the cache (Section~\ref{sec:caching}) before reporting the outcome.

A writable operation cannot be left open to every caller. On the client side, \textit{REQ8} is handled through the \texttt{\#auth required} field. The field can be set per operation or as an API-wide default; operation-level settings override the default. It places the operation behind bearer-token authentication~\cite{jones_oauth_2012}: the caller presents a token in the HTTP \texttt{Authorization} header with the \texttt{Bearer} scheme, and a request carrying no token, or one that is unknown, expired, or revoked, is rejected with an HTTP~401 response. Whatever is left unmarked stays reachable without credentials, so read access is undisturbed. The tokens live in a local SQLite registry that keeps only the SHA-256 hash of each one. The command line creates a token with an optional label and lifetime, lists the active ones, and revokes any of them.

\textit{REQ9} moves to the boundary between RAMOSE and the database: the triplestore itself may require authentication, and RAMOSE must present a credential to query it. A credential is registered per endpoint, through a command-line option or, preferably for secrets, an environment variable, as the full value of an \texttt{Authorization} header. RAMOSE replays that header on every request it sends to the matching endpoint, reads and writes alike, and to that endpoint only. Because the value travels unchanged, the scheme behind it is irrelevant, be it HTTP Basic, a bearer token, or a vendor-specific one. HTTP Digest is the exception, since the store issues a new challenge that each answer must match: for such an endpoint the credential is registered as a user and a password, and RAMOSE computes the answer to every challenge.

\section{Community's uptake and use}
\label{sec:uptake}

This section describes how RAMOSE has been adopted and reused in several contexts: OpenCitations, SKG-IF, and GRAPHIA.

\subsection{OpenCitations}
\label{sec:opencitations}

\begin{figure}[!htb]
  \centering
  \includegraphics[width=\columnwidth]{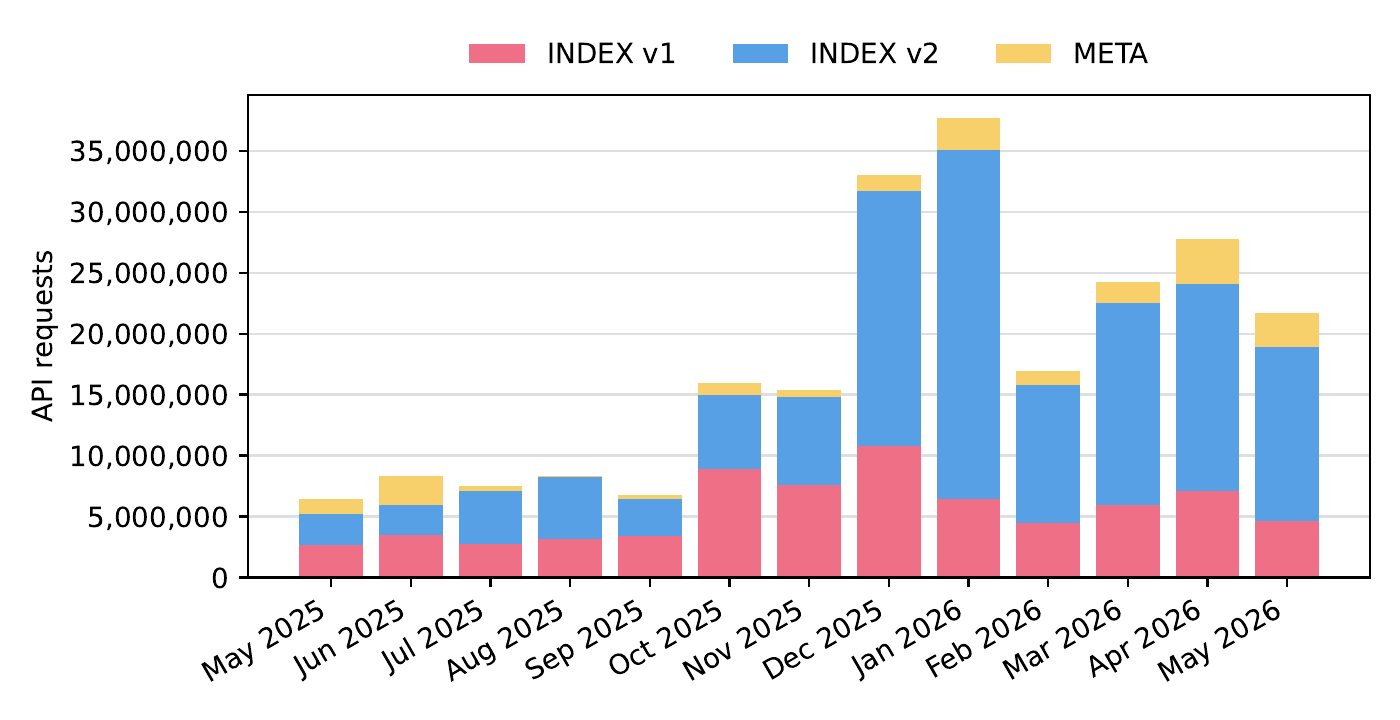}
  \def\xfigwd{0pt}
  \caption{Monthly requests to the OpenCitations REST APIs between May 2025 and May 2026, divided by API: INDEX v1, INDEX v2, and META. Data from \url{https://statistics.opencitations.net/}.}
  \label{fig:oc_stats_1}
\end{figure}

The OpenCitations statistics page (\url{https://statistics.opencitations.net/}) monitors the use of the OpenCitations services, whose REST APIs are all served through RAMOSE. Figure~\ref{fig:oc_stats_1} reproduces its data for the twelve months between May 2025 and May 2026, a period in which the OpenCitations APIs were still run by the first version of RAMOSE. It traces the monthly number of API requests, broken down by API: INDEX v1, INDEX v2, and META. The volume ranges from about 6.4 million requests in May 2025 to a peak of almost 38 million in January 2026, with the two INDEX APIs accounting for most of the traffic. The raw access logs for January to April 2026 are published on Zenodo~\cite{petrella_opencitations_2026}. As a term of comparison, the article presenting the first version of RAMOSE~\cite{daquino_creating_2022} reported 4,394,093 API calls in the first quarter of 2020; in the first quarter of 2026, the OpenCitations APIs received 79,369,678, roughly eighteen times as many. These figures therefore show the demand that the new version inherits, while its first deployment is the SKG-IF endpoint described in Section~\ref{sec:graphia}.

\subsection{SKG-IF and GRAPHIA}
\label{sec:graphia}

In the context of the GRAPHIA Project (\url{https://cordis.europa.eu/project/id/101188018/}), SKG-IF has been adopted as the basic interoperability framework to federate multiple systems, thereby exposing their data as part of a giant information graph, with SKG-IF serving as the data model and format. The entry point that a new source must provide to be included in such a federation is, indeed, to expose an instance of the SKG-IF REST API. However, developing a new REST API from scratch by mapping the actual data model of a source into the result format exposed by SKG-IF is not a trivial task at all, since it costs several person-hours to implement and maintain, even in the cases in which the original source exposes its data using Semantic Web technologies (e.g. via a SPARQL endpoint).

In this context, RAMOSE, and in particular its new feature described in Section~\ref{sec:skgif}, has been adopted by the GRAPHIA project for two (short-term and long-term) purposes. On the one hand, RAMOSE can be used by any of the sources already involved in the project, particularly those already exposing their data via a SPARQL endpoint, to simplify the creation of a SKG-IF REST API, thus speeding up the onboarding process of current sources into the GRAPHIA Federation -- as recently done by OpenCitations (\url{https://api.opencitations.net/skg-if/v1}), one of the infrastructures involved in GRAPHIA. On the other hand, GRAPHIA also wants to develop a specific module to enable future sources, external to the project, to be easily included in the GRAPHIA Federation, thereby increasing the number of relevant data providers -- and RAMOSE has been suggested as one of the primary candidates to fulfil this purpose.

\section{Related works and functional comparison}
\label{sec:related}

Several tools share RAMOSE's goal of exposing RDF data through REST APIs. Their principal difference lies in the kind of input each one expects from the developer who configures the API. A first group requires the query that each operation runs: grlc~\cite{merono-penuela_grlc_2016,merono-penuela_grlc_2026}, BASIL~\cite{daga_basilar_2015, daga_basil_2025}, R4R~\cite{badenes-olmedo_r4r_2021, badenes-olmedo_r4r_2019}, and RDFProxy~\cite{plank_declarative_2025, schacht_rdfproxy_2026} take SPARQL, as does RAMOSE itself, while Walder~\cite{heyvaert_walder_2023} takes a query written in GraphQL-LD~\cite{taelman_graphqlld_2018} or in SPARQL. A second group requires a higher-level description from which the engine generates the SPARQL: CRAFTS~\cite{vega-gorgojo_crafts_2022} from a JSON description of the resources and their properties, OBA~\cite{garijo_oba_2020, osorio_oba_2022} from an OWL ontology, Elda~\cite{epimorphics_elda_2026} from a Linked Data API (LDA) description~\cite{reynolds_linked_2015}, and ShExpose~\cite{gopfert_shexpose_2026} from ShEx shape expressions.

Most of these tools have already been compared by Espinoza-Arias \textit{et al.}~\cite{espinoza-arias_crossing_2021}, whose survey evaluates them along fourteen criteria: Year, Interface Description Language, Input, Output, Operations, Configuration format, Configurable queries, Authentication, Resources, Versioning, Control over the JSON structure, Source, Last release, and Language. We keep from that survey the tools comparable with RAMOSE, i.e. the server-side generators of REST APIs over SPARQL endpoints, reuse its criteria, and add CRAFTS, RDFProxy, and ShExpose, which appeared after it. For clarity, we split the criteria across two tables: Table~\ref{tab:profile} gathers the four metadata criteria of the survey (Year, Language, Source, and Last release) with the addition of the license under which each tool is released, while Table~\ref{tab:comparison} compares the tools with RAMOSE along the functional criteria. To what the survey already covered, we add five dimensions central to RAMOSE v2 given the requirements of Section~\ref{sec:requirements}: querying multiple endpoints, joining results across independent queries, reading non-RDF sources, pagination, and caching. We also split the survey's single Authentication criterion, which only weighed whether the generated API asks its clients for credentials, into a consumer-side and an endpoint-side row. RAMOSE appears as separate v1 and v2 columns to mark what changed since the 2022 paper.

\begin{table*}[!htb]
\caption{Descriptive metadata of the compared tools. Year is the repository creation year; ``---'' marks an absent value.}
\label{tab:profile}
\centering
\footnotesize
\setlength{\tabcolsep}{6pt}
\begin{tabular}{l c l l l l}
\toprule
Tool & Year & Language & License & Source & Last release \\
\midrule
RAMOSE & 2018 & Python & ISC & \texttt{opencitations/ramose}~\cite{massari_ramose_2026} & v2.10.0 (2026) \\
grlc & 2015 & Python & MIT & \texttt{CLARIAH/grlc}~\cite{merono-penuela_grlc_2026} & v1.3.11 (2026) \\
BASIL & 2015 & Java & Apache 2.0 & \texttt{basilapi/basil}~\cite{daga_basil_2025} & v1.0.0-RC1 (2025) \\
OBA & 2019 & Java & Apache 2.0 & \texttt{KnowledgeCaptureAndDiscovery/OBA}~\cite{osorio_oba_2022} & 3.6.1 (2022) \\
R4R & 2019 & Java & Apache 2.0 & \texttt{cbadenes/r4r}~\cite{badenes-olmedo_r4r_2019} & 0.4 (2019) \\
CRAFTS & 2021 & JavaScript & Apache 2.0 & \texttt{guiveg/crafts}~\cite{vega-gorgojo_crafts_2022a} & --- (2022) \\
RDFProxy & 2024 & Python & GPLv3 & \texttt{acdh-oeaw/rdfproxy}~\cite{schacht_rdfproxy_2026} & v0.11.0 (2026) \\
Elda & 2013 & Java & Apache 2.0 & \texttt{epimorphics/elda}~\cite{epimorphics_elda_2026} & 3.0.4 (2026) \\
Walder & 2020 & JavaScript & MIT & \texttt{KNowledgeOnWebScale/walder}~\cite{heyvaert_walder_2023} & v4.1.4 (2023) \\
ShExpose & 2026 & TypeScript & --- & \texttt{chgoe/ShExpose-REST-API-Generation-from-Shapes}~\cite{gopfert_shexpose_2026a} & --- (2026) \\
\bottomrule
\end{tabular}
\end{table*}

\begin{table*}[!htb]
\caption{Functional comparison of server-side REST API generators over SPARQL endpoints, reusing the criteria of Espinoza-Arias \textit{et al.}~\cite{espinoza-arias_crossing_2021} (top eleven rows) and adding five dimensions specific to RAMOSE v2 (bottom five). \yes{} supported, \no{} not supported, \pmark{} partial. For Resources, S, M, and N denote a single resource, multiple resources (a flat collection), and nested resources (a resource embedding a subcollection of related resources).}
\label{tab:comparison}
\centering
\scriptsize
\setlength{\tabcolsep}{2.5pt}
\renewcommand{\arraystretch}{1.15}
\begin{tabular}{>{\raggedright\arraybackslash}p{2.2cm} *{11}{>{\centering\arraybackslash}p{1.05cm}}}
\toprule
Dimension & RAMOSE v1 & RAMOSE v2 & grlc & BASIL & OBA & R4R & CRAFTS & RDFProxy & Elda & Walder & ShExpose \\
\midrule
Interface Description Language & HTML & OpenAPI 3.2, HTML & OpenAPI 2.0 & Swagger 1.2 & OpenAPI 3.0 & --- & OpenAPI 3.0 & OpenAPI 3.1 & LDA spec & OpenAPI 3.0 & OpenAPI 3.0 \\
Input & SPARQL & SPARQL & SPARQL & SPARQL & OWL ontology & SPARQL, templates & JSON config & SPARQL, model & RDF spec & GraphQL-LD, SPARQL & ShEx \\
Output & CSV, JSON & Any & endpoint-dependent & XML, JSON, CSV, RDF & JSON & JSON & JSON & JSON & Any & HTML, JSON, RDF & JSON \\
Operations & GET & CRUD & GET, POST & GET, POST & CRUD & GET & CRUD, PATCH & GET & GET & GET & CRUD \\
Configuration format & \texttt{.hf} & \texttt{.hf}, YAML & \texttt{.rq}, YAML & REST API & YAML & \texttt{.sparql}, \texttt{.vm} & JSON & Pydantic model & RDF/Turtle & YAML & LinkML YAML \\
Configurable queries & \yes & \yes & \yes & \yes & \yes & \yes & \yes & \yes & \yes & \yes & \no \\
Consumer auth. & \no & Bearer & \no & \no & Bearer & Basic & Basic, Bearer & \no & \no & \no & \no \\
Endpoint auth. & \no & Basic, Bearer, Digest & Basic & Basic & Basic & \no & Basic, Digest & Basic, Digest & Basic & \no & Basic, token \\
Resources & S, M & S, M, N & S, M, N & S, M & S, M, N & S, M, N & S, M, N & S, M, N & S, M, N & S, M, N & S, N \\
Versioning & \yes & \yes & \yes & \no & \yes & \no & \no & \no & \no & \no & \no \\
Control over JSON & \yes & \yes & \yes & \no & \no & \yes & \yes & \yes & \yes & \yes & \no \\
\midrule
Multiple endpoints & \no & \yes & \yes & \no & \no & \no & \yes & \no & \no & \yes & \no \\
Non-RDF sources & \no & \yes & \no & \no & \no & \no & \no & \no & \no & \no & \no \\
Join across queries & \no & \yes & \no & \no & \no & \no & \pmark & \no & \no & \pmark & \no \\
Pagination & \no & \yes & \yes & \no & \pmark & \pmark & \no & \yes & \yes & \pmark & \no \\
Caching & \no & \yes & \no & \no & \no & \no & \yes & \no & \yes & \no & \no \\
\bottomrule
\end{tabular}
\end{table*}

All the cells of Table~\ref{tab:comparison} were checked by running the tools against the same lookup: a single bibliographic resource retrieved by DOI, whose metadata and citations live in two separate stores. The stores are two local snapshots of OpenCitations Meta and OpenCitations Index, reduced to the triples the lookup needs and served by Apache Jena Fuseki, so that the results do not depend on the version of the public endpoints. These tests are published as part of RAMOSE's documentation (\url{https://opencitations.github.io/ramose/comparison/intro.html}) as a set of Jupyter notebooks.

The paragraphs that follow describe the tools, first those that ask the developer to write the query and then those that generate it from a higher-level description. \textbf{grlc} builds an API at request time because it treats the SPARQL queries as first-class artefacts, versioned in their own repository: each query file becomes one operation, and since the queries are read when the request arrives, editing one changes the API without redeploying the server. Behaviour beyond the bare query is declared in comments placed above it. These add the endpoint, pagination, and a transformation that reshapes the JSON result. grlc generates an OpenAPI 2.0 description whose version field is the git commit the queries were read from. By naming a commit in the request URL, a client can call the API as it was at any past revision of the query repository, and the description links the neighbouring commits. For output, grlc forwards the request's \texttt{Accept} header to the endpoint and returns whatever serialisation the endpoint supports.

\textbf{BASIL} also starts from a SPARQL query, but stores it in its own store (Jena TDB2 by default) via \texttt{PUT} requests. These calls can be made directly or issued through PESTO~\cite{daga_pesto_2016}, a web GUI. The response format follows from HTTP content negotiation or from a file extension on the path (\texttt{.json}, \texttt{.ttl}); unlike grlc, BASIL does not defer serialisation to the endpoint but renders the results itself, so the same fixed set of formats, namely JSON, XML, CSV, and RDF serialisations, is available whatever the endpoint emits. The XML is a flat table of the result rows in BASIL's own layout or the W3C SPARQL Query Results XML Format~\cite{beckett_sparql_2013}. HTTP Basic authentication protects only the operations that create or modify an API, while calling one is always open.

\textbf{R4R} derives its API from a directory tree: each folder under the resource root is a resource, the SPARQL query files it contains are the operations on that resource, and a subfolder is a sub-resource under the parent's path. R4R does not serialise results itself but passes the rows to a developer-written Apache Velocity template. This should in principle allow any output format, yet the tool is wired for JSON alone: the Content-Type is hard-coded to \texttt{application/json} and, as we verified with Turtle and CSV templates, R4R JSON-escapes every value, so IRIs, DOIs, and URLs are emitted as valid JSON strings but malformed Turtle and CSV. The query files carry no endpoint, so a single SPARQL endpoint is bound for the whole server. HTTP Basic authentication is available only globally, guarding reads and writes, and the tool produces no API description.

\textbf{RDFProxy} maps a SPARQL result onto a Pydantic model that validates the response, and FastAPI generates an OpenAPI 3.1 description from that model. Its pagination is entity-aware: rather than slicing the raw result rows, it rewrites the query to paginate by distinct entities, so the rows of one nested entity remain on the same page. A single adapter binds one endpoint, so RDFProxy cannot combine two sources.

\textbf{Walder} takes a YAML configuration written in the OpenAPI 3.0 structure, where each route binds a query to a list of data sources and view templates. The query is written either in GraphQL-LD or as a raw SPARQL \texttt{CONSTRUCT}. Querying and federation are delegated to Comunica~\cite{taelman_comunica_2018}, so a source can be a Solid pod~\cite{sambra_solid_2016}, a SPARQL endpoint, a Triple Pattern Fragments interface~\cite{verborgh_triple_2016}, or an RDF file. The developer only lists the sources, without stating which predicate lives where: Comunica sends each query pattern to every source and joins the partial results on their shared variables.

Rather than one SPARQL query per operation, \textbf{CRAFTS} asks the developer to write a JSON configuration that describes the data: it lists the types of resource and their properties, and for each property it states from which endpoint and with which RDF predicate the value is read. From any such configuration, CRAFTS exposes one uniform REST interface: an operation to fetch a resource by its IRI, one to fetch several at once, and one to run a configured query template. The join itself rests on this per-property binding: because each property names its own endpoint, one resource request combines a resource's properties from several triple stores. However, the join is keyed by IRI: the same IRI must identify the entity in every store, and the client must supply it; the request involving a join cannot start from any other value. A further feature concerns nested resources: when a resource points to another through an object property, CRAFTS can inline it, nesting its data in the JSON instead of returning only its IRI. It also caches results and supports write operations, including partial updates through HTTP PATCH.

\textbf{OBA} derives its API from an OWL ontology. Each class becomes two operations, one for the collection and one for a single instance addressed by its identifier. At request time, OBA fills a \texttt{CONSTRUCT} template, takes the JSON-LD the endpoint returns, frames it against the generated context and removes the \texttt{@context}, leaving the client with plain JSON; the same specification validates each request. Write access (POST, PUT, DELETE) requires a bearer token that the server issues, while reads stay open.

\textbf{Elda} likewise generates the SPARQL from a declarative description, but one of the API rather than of the data. The developer writes the API as an RDF document in the Linked Data API vocabulary: each operation declares a graph pattern that picks the matching resources and a list of property paths, such as \texttt{identifier.value} for a record's DOI, that name the fields to return. From this, Elda generates two SPARQL queries, one to find the resources and one to fetch those fields, and the short label attached to each property serves at once as a request parameter and as a key in the returned JSON. The viewer that fetches those fields is always generated from the property list, so Elda returns stored values rather than computed ones, and because each API binds a single endpoint, joins across independent sources cannot be expressed. In return, the model supplies, with no extra configuration, mandatory paging whose navigation travels as RDF metadata (not as RFC 8288 \texttt{Link} headers), pluggable formatters for JSON, JSON-LD, RDF/XML, Turtle, XML, HTML, and Atom, and a per-endpoint cache. Elda reads only, and its authentication protects the upstream endpoint, not the API's consumers.

In \textbf{ShExpose}~\cite{gopfert_shexpose_2026} the developer uses LinkML~\cite{moxon_linked_2021}, a YAML schema language, to model the data, and a generator distributed with LinkML translates that model into a ShEx schema~\cite{prudhommeaux_shape_2014}; from each shape, ShExpose derives a set of CRUD routes together with the SPARQL that reads and writes the matching subgraph. With this approach, a resource can be retrieved only by its subject IRI, not by a property value. ShExpose binds a single endpoint, so it cannot join across sources.

Finally, CRAFTS, OBA, and Elda let developers write their own SPARQL query when their declarative description is not expressive enough, while ShExpose does not.

Most dimensions in Table~\ref{tab:comparison} are self-explanatory; resources, multiple endpoints, joining, and pagination rest on classification choices that warrant explanation.

Resources encode three cases. Following Espinoza-Arias \textit{et al.}, \emph{S} marks a tool that exposes a single resource, \emph{M} a flat collection of resources, and \emph{N} nested resources, that is, a resource that embeds a subcollection of related resources. For instance, RAMOSE v2 produces this nesting through its pluggable format system, of which the SKG-IF module is one instance, and grlc through its \texttt{transform} comment, both grouping sibling result rows under a shared key. RAMOSE v1 serialised the result to CSV before building the JSON, so the output was always a flat list of records, and no custom post-processing function could alter that shape. This is why it is marked S, M rather than S, M, N.

Multiple endpoints are supported when a single API reaches more than one endpoint. One RAMOSE v1 instance could serve several APIs, but each of them was bound to its own endpoint, so v1 is marked as unsupported. Joining across sources also warrants a note. Among the tools reaching more than one endpoint, CRAFTS and Walder join only on terms that already coincide in the data, on a shared IRI, and on shared query variables, respectively, and only within a single request; the others bind to one endpoint. RAMOSE v2 instead splits an operation into steps driven by directives and joins the resulting tables on an arbitrary key.

We count pagination as supported only when three elements appear together: a request parameter that selects a bounded window of the result, a navigation system to adjacent pages, and a termination signal that lets a client know when the result is exhausted. A tool that injects only \texttt{LIMIT} and \texttt{OFFSET} in the SPARQL query offers windowing rather than pagination, which we mark as partial. By this measure, RAMOSE v2, grlc, RDFProxy, and Elda paginate fully; OBA, R4R, and Walder expose only the windowing parameter; and BASIL and CRAFTS provide none of the three. Where the window is applied also varies: RAMOSE's generic pipeline paginates the result rows, but a pluggable formatter can make it entity-aware, as the SKG-IF module does by grouping the rows into entities and paginating those, the same boundary-respecting behaviour RDFProxy achieves by rewriting the query so that an entity with several rows is never split across a page.

Across all systems compared, RAMOSE v2 is the only one that can query RDF and non-RDF sources in one API operation and join their results.

\section{Performance evaluation}
\label{sec:evaluation}

This section measures the multi-source directives of Section~\ref{sec:multisource} against a single SPARQL query with a \texttt{SERVICE} clause. Both strategies answer the same request: given the ISSN of a venue, return its articles, their incoming citations, and the metadata of the citing works. The articles and the metadata live in OpenCitations Meta, while the citations live in OpenCitations Index. In the federated strategy, the Meta store resolves the ISSN and the articles, delegates the citations to the Index endpoint through \texttt{SERVICE}, and enriches the citing works locally. The orchestrated strategy is a longer version of the operation of Listing~\ref{lst:multisource}, with the same structure. Its last query also returns the identifiers, authors, and venue of each citing work. RAMOSE runs one query on Meta for the articles and one query on Index with the articles injected through \texttt{@@values}. It then runs one query on Meta for every batch of 3,000 citing works and joins the rows between the steps. Since the two operations live in the same specification file, they share the endpoint setup and differ only in the query text.

The Meta snapshot of 27 June 2026, with about 5.1 billion triples~\cite{opencitations_opencitations_2026}, is loaded in Virtuoso, which acts as the federating engine. The Index snapshot, with about 10.2 billion triples~\cite{opencitations_opencitations_2026a}, is loaded in QLever, which acts as the remote endpoint. Both stores run in Docker on one host with an Intel Core i9-12900K, which has 8 performance cores and 8 efficient cores, and 125 GB of memory. Virtuoso receives 64 GB of memory and QLever 16 GB. These values were chosen empirically by raising them until the federated query completed every request without a timeout and without running out of resources. The stores are therefore not the bottleneck, and the federated query is measured in its best case. The load stops at eight concurrent clients for the same reason, because with one client per performance core the host does not limit the service. An HTTP proxy in front of each store records every request with its size and latency: this allowed us to count the exact number of federated requests that Meta sends to Index. RAMOSE serves the operations through gunicorn, and every layer grants 600 seconds to a request. The sample holds 20 ISSNs drawn from Meta, five for each band of published articles: 1-9, 10-99, 100-999, and 1,000-9,999. The largest venue has 9,427 articles, while the venue with most citing works has 85,291. Before the timed runs, every response of both strategies is checked against a reference answer: a faster strategy cannot win by returning fewer rows. Each ISSN is then requested three times per strategy, first with one client and then with eight. The metrics are the backend requests behind each API call, reported as mean and maximum per band, and the end-to-end latency, reported as median and 95th percentile. Table~\ref{tab:federation} shows the outcome. Every request returned HTTP 200 with the expected rows.

\begin{table*}[!htb]
\caption{Latency and backend requests of the federated \texttt{SERVICE} query and of the RAMOSE orchestration for the same venue request. Each row summarises 15 API calls: five venues with the given number of articles, each requested three times, with one client or with eight clients at once. The median and 95th percentile are end-to-end latencies of one API call in milliseconds. Requests are the calls that the proxies recorded behind one API call, given as the mean and the maximum over 15 calls. They are counted with one client, where the calls are serial, and every backend request belongs to a known call.}
\label{tab:federation}
\centering
\footnotesize
\setlength{\tabcolsep}{6pt}
\begin{tabular}{l r r r r r r r}
\toprule
& & \multicolumn{3}{c}{\texttt{SERVICE}} & \multicolumn{3}{c}{Orchestration} \\
\cmidrule(lr){3-5} \cmidrule(lr){6-8}
Articles & Clients & Median (ms) & 95th pct. (ms) & Requests (mean / max) & Median (ms) & 95th pct. (ms) & Requests (mean / max) \\
\midrule
1--9 & 1 & 156.8 & 334.9 & 5.2 / 9 & 64.3 & 81.2 & 3 / 3 \\
1--9 & 8 & 174.3 & 363.2 & 5.2 / 9 & 85.7 & 129.0 & 3 / 3 \\
10--99 & 1 & 3,029.5 & 3,564.1 & 66.6 / 98 & 118.1 & 524.0 & 3 / 3 \\
10--99 & 8 & 3,018.0 & 3,634.8 & 66.6 / 98 & 239.9 & 1,437.6 & 3 / 3 \\
100--999 & 1 & 12,557.6 & 14,042.1 & 246.8 / 301 & 215.6 & 290.8 & 3 / 3 \\
100--999 & 8 & 12,492.9 & 14,254.7 & 246.8 / 301 & 435.6 & 699.3 & 3 / 3 \\
1,000--9,999 & 1 & 207,908.3 & 414,146.6 & 5,605.2 / 9,428 & 10,319.4 & 29,526.5 & 16.6 / 31 \\
1,000--9,999 & 8 & 211,291.5 & 413,790.2 & 5,605.2 / 9,428 & 13,063.5 & 40,827.3 & 16.6 / 31 \\
\bottomrule
\end{tabular}
\end{table*}

The number of remote requests explains the gap. Virtuoso sends the Index endpoint one request for every article, so the largest venue needs 9,427 remote requests. The orchestration needs at most 31 requests across both stores, because it packs thousands of articles or citing works in each request. Each remote request costs a median of about 11 ms. The federated query issues them one after the other and waits for each answer before it sends the next one, so 9,427 short waits add up to minutes. The five venues of the largest band take from 61 to 413 seconds with a single client, against 0.6 to 29 seconds for the orchestration. The smallest band confirms this reading: with one client, where the remote work is at its lowest, the gap almost vanishes, and the orchestration is only about twice as fast as the federated query. In the other bands, where the remote work dominates, it is from 20 to 58 times faster. Eight clients change the picture in a way that deserves a comment. The federated latency stays within one percent of the single-client value. A federated request spends almost all its time waiting for the remote answers, one at a time, and the engines are almost idle in the meantime, so eight such requests do not disturb each other. The orchestration spends its time differently: its few requests carry thousands of items each, and the engines must compute on them. Eight of these requests arrive at once and share the same cores, so the latency of the orchestration grows with the load. Even so, it asks the engines for far less work in total, and it remains from 2 to 29 times faster than the federated query.

These results depend on the use of Virtuoso as the federating engine and of QLever as the remote engine, and on the type of query. Another federating engine may group the bindings in another way and send fewer remote requests. Orchestration removes this dependence. With a federated query, the engine that serves the query decides the number and shape of the remote requests, and the API author has no say in them. With orchestration, the same choices are written in the specification file: the API author sets the batch size and the order of the steps, and the operation works the same way regardless of what store sits behind each source. The benchmark is part of the RAMOSE repository~\cite{massari_ramose_2026} and can be reproduced with a single shell command.

\section{Discussion and conclusions}
\label{sec:conclusions}

Several technical limitations remain. Two concerns relate to the write operations introduced in Section~\ref{sec:auth}. First, an operation that modifies data targets a single endpoint: the multi-source directives of Section~\ref{sec:multisource} orchestrate reads across several sources, but they do not extend to updates, so a write cannot act on more than one endpoint at a time. Second, a successful write clears the whole cache rather than invalidating only the entries the update affects.

A third limitation concerns the cache and the token store, both backed by local SQLite files. The present design does not scale horizontally, and although SQLite tolerates concurrent access, its locking model degrades performance under parallel load.

On the access-control side, authorisation is binary at the granularity of an operation: a valid token grants access to every operation marked as requiring authentication.

Finally, RAMOSE reaches non-RDF sources exclusively through SPARQL Anything, which can query file formats such as CSV, JSON, and XML but not relational databases, limiting applicability at scale.

These limitations point to concrete directions for future work. Extending the multi-source directives to update operations would let a write act across several endpoints, but it raises problems the read-only orchestration does not face. The foremost is atomicity: SPARQL endpoints provide no cross-endpoint transaction protocol, so a failure midway would leave some sources updated and others not, and RAMOSE would need its own coordination, whether by idempotent retries or a best-effort model that reports partial outcomes to the client. The orchestration would also have to interleave reads and writes, since a write step often depends on a value read from another source.

The cache could be cleared selectively by analysing the SPARQL \texttt{UPDATE} in a write to identify the entities it changes, removing only the cache entries that involve those entities. Authorisation could be made finer-grained by introducing scopes, roles, or per-token permissions. Finally, a pluggable cache backend would let a provider replace the local SQLite store with an external store such as Redis, lifting the horizontal-scaling and concurrency limits.

Alongside these developments, we also plan to broaden the range of standards RAMOSE supports. The pluggable module that packages SKG-IF support (Section~\ref{sec:skgif}) is, at present, the only built-in instance of its kind; extending the same mechanism to additional community interoperability standards is a natural next step, and the Open Archives Initiative Protocol for Metadata Harvesting~\cite{lagoze_open_2015} is one candidate we intend to explore.

Section~\ref{sec:evaluation} measures the RAMOSE orchestration against a federated query, but it compares two execution strategies within one tool; a benchmark across tools remains to be done. Existing benchmarks for RDF and SPARQL systems, such as LargeRDFBench~\cite{saleem_largerdfbench_2018} and the Berlin SPARQL Benchmark~\cite{bizer_berlin_2009}, assess triplestores and query engines, i.e. the layer beneath RAMOSE, while the comparison on which Section~\ref{sec:related} builds~\cite{espinoza-arias_crossing_2021} was functional rather than performance-oriented. To the best of our knowledge, no benchmark currently targets the middleware layer of REST-API generators over SPARQL endpoints. Therefore, we plan to develop a dedicated benchmark that fixes several endpoints, so that both single-source and federated queries are exercised against a shared data model. OpenCitations Meta and OpenCitations Index datasets, large-scale RDF databases holding approximately 5.1 billion and 10.2 billion triples, respectively, as of June 2026, are possible candidates for these endpoints. Each tool will run against them with the query configuration its authors consider optimal, developed in collaboration with them, so differences in measured latency and throughput reflect the tools' architectures rather than a single party's tuning effort.

RAMOSE is available on PyPI under the ISC license, together with its source code, documentation, and the reproducible comparison of Section~\ref{sec:related}~\cite{massari_ramose_2026}. By reducing the effort needed to expose data through documented REST APIs that follow shared specifications, RAMOSE aims to make joining interoperability networks such as SKG-IF a matter of configuration rather than development, thereby widening the network of providers that expose their data in an interoperable way.

\section*{Declaration of generative AI and AI-assisted technologies}

The authors declare that no generative AI was used in the writing of this manuscript.

\section*{Declarations}
\subsection*{Authors' contribution statements}
\begin{itemize}
\raggedright
    \item \textbf{Arcangelo Massari:} Conceptualization, Investigation, Methodology, Resources, Software, Validation, Visualization, Writing -- original draft, Writing -- review \& editing.
    \item \textbf{Sergei Slinkin:} Conceptualization, Investigation, Methodology, Resources, Software, Writing -- original draft.
    \item \textbf{Ivan Heibi:} Conceptualization, Methodology, Software, Supervision, Writing -- review \& editing.
    \item \textbf{Silvio Peroni:} Conceptualization, Funding acquisition, Methodology, Resources, Software, Supervision, Writing -- original draft, Writing -- review \& editing.
\end{itemize}

\bibliographystyle{IEEEtran}
\bibliography{references}

\begin{IEEEbiographynophoto}{Arcangelo Massari} (\url{https://orcid.org/0000-0002-8420-0696}) received the joint Ph.D. degree in Cultural Heritage in the Digital Ecosystem from the University of Bologna and in Engineering Technology from KU Leuven in 2026.

He is currently a Postdoctoral Researcher at the University of Bologna, working on the GRAPHIA project (Horizon Europe), which aims to integrate fragmented data across the Social Sciences and Humanities into a single knowledge graph. He also collaborates with OpenCitations, where he works on maintaining the bibliographic metadata of OpenCitations Meta and on developing interfaces that enable non-technical domain experts to curate cultural heritage data with integrated provenance and change tracking.
\end{IEEEbiographynophoto}

\begin{IEEEbiographynophoto}{Sergei Slinkin} (\url{https://orcid.org/0009-0002-6830-7765}) received the master's degree in Digital Humanities and Digital Knowledge from the University of Bologna in 2026, with a dissertation on extending RAMOSE to multiple output formats and distributed data sources.
\end{IEEEbiographynophoto}

\begin{IEEEbiographynophoto}{Ivan Heibi} (\url{https://orcid.org/0000-0001-5366-5194}) received the Ph.D. degree in computer science from the University of Bologna, where his doctoral research focused on applying Semantic Web technologies to the Science of Science domain, particularly in the Arts and Humanities.

He is currently an Assistant Professor at the University of Bologna, Department of Classical Philology and Italian Studies, working primarily on the PE5 – CHANGES project and involved in Virtual Technologies for Museums and Art Collections. He has contributed to the development and management of the OpenCitations infrastructure, where he is the Chief Technology Officer. His publications include articles on open citation data and semantic artefacts in international venues. His research interests include Semantic Web technologies, Information Science, bibliometrics, data visualisation, and automatic document analysis. 

Prof. Heibi is a member of the Research Centre for Open Scholarly Metadata and the Digital Humanities Advanced Research Centre.
\end{IEEEbiographynophoto}

\begin{IEEEbiographynophoto}{Silvio Peroni} (\url{https://orcid.org/0000-0003-0530-4305}) holds a PhD degree in Computer Science and is an Associate Professor at the Department of Classical Philology and Italian Studies, University of Bologna. He is Director of the Research Centre for Open Scholarly Metadata, one of the main developers of the SPAR Ontologies and Director of OpenCitations. He is an Editorial Board member of Journal of Documentation and PeerJ Computer Science, among others, and he has been involved in the organisation and programme committee of several Semantic Web venues. His works concern theoretical studies and technical implementation of tools to foster semantic interoperability of Open Science services and infrastructures, the empirical analysis of the nature of scholarly citations, bibliometrics and scientometrics studies, visualisation and browsing interfaces for semantic data, and the development of ontologies to manage, integrate and query bibliographic information and cultural heritage data.
\end{IEEEbiographynophoto}

\EOD
\vspace*{\fill}

\end{document}